\documentclass[preprint] {article}
\usepackage[utf8]{inputenc} % set input encoding (not needed with XeLaTeX)
\usepackage{amsmath}
\usepackage{amsfonts}
\usepackage{amssymb}
\usepackage{geometry}
\allowdisplaybreaks
\newcommand{\be}{\begin{equation}} \newcommand{\ee}{\end{equation}}
\newcommand{\bea}{\begin{eqnarray}}\newcommand{\eea}{\end{eqnarray}}

\newcommand{\g}{\nabla}

\usepackage{geometry} % to change the page dimensions
\begin{document}
\title{On Symmetries and Exact Solutions of a Class of Non-local Non-linear
Schr$\ddot{o}$dinger Equations with Self-induced ${\cal{PT}}$-symmetric
Potential
}
\author{Debdeep Sinha\footnote{{\bf email:}  debdeepsinha.rs@visva-bharati.ac.in} 
\ and Pijush K. Ghosh \footnote {{\bf email:}
pijushkanti.ghosh@visva-bharati.ac.in}}
\date{Department of Physics, Siksha-Bhavana, \\ 
Visva-Bharati University, \\
Santiniketan, PIN 731 235, India.}
\maketitle
\begin{abstract}
A class of non-local non-linear Schr$\ddot{o}$dinger equations(NLSE)
is considered in an external potential with space-time modulated coefficient
of the nonlinear interaction term as well as confining and/or loss-gain
terms.  This is a generalization of a recently introduced integrable non-local
NLSE with self-induced potential that is ${\cal{PT}}$ symmetric in the
corresponding stationary problem. Exact soliton solutions are obtained for the
inhomogeneous and/or non-autonomous non-local NLSE by using similarity
transformation and the method is illustrated with a few examples. It is found
that only those transformations are allowed for which the transformed spatial
coordinate is odd under the parity transformation of the original one. It is
shown that the non-local NLSE without the external potential and a $d+1$
dimensional generalization of it, admits all the symmetries of the $d+1$
dimensional Schr$\ddot{o}$dinger group. The conserved Noether charges
associated with the time-translation, dilatation and special conformal
transformation are shown to be real-valued in spite of being non-hermitian.
Finally, dynamics of different moments are studied
with an exact description of the time-evolution of the ``pseudo-width'' of
the wave-packet for the special case when the system admits a $O(2,1)$
conformal symmetry.
\end{abstract}
\tableofcontents
\vspace{0.3in}

%introduction

\section{Introduction}
Ever since it was realized that $\cal{PT}$ symmetric non-hermitian systems
may exhibit real spectra\cite{aa}, a great deal of investigation has been
carried out in this field\cite{ali,znojil,piju,fring,longhi,fabio}. 
As theoretical understanding proceeds,
attempts have been made to realize non-hermitian $\cal{PT}$ symmetric
systems experimentally. Since, the paraxial equation of diffraction is similar
in structure to the Schrodinger equation, it was believed that optics may be
a testing ground for $\cal{PT}$ symmetric systems\cite{kg}. In fact, the phase
transition between broken and unbroken phases of a non-hermitian system has
been observed experimentally\cite{cer}, stimulating a great deal of
research\cite{r1, r10} in optical systems with ${\cal{PT}}$ symmetry. 

The non-linear Schrodinger equation(NLSE) admits soliton solutions and finds
applications in many diverse branches of modern science like Bose-Einstein
condensation\cite{r2}, plasma physics\cite{r3}, gravity waves\cite{r4},
$\alpha$-helix protein dynamics\cite{r5} etc. and specially optics where it\
describes wave propagation
in non-linear media\cite{ac}. The study of soliton in NLSE was mainly confined
to homogeneous and autonomous systems during the earlier years of its
development, where time merely played the role of a parameter in the nonlinear
evolution equation. However, it became apparent that integrability of NLSE may
be preserved\cite{cl}, if different co-efficients appearing in it are given
specific space-time dependencies. This lead to the concept of non-autonomous
solitons\cite{vn}. A great deal of research\cite{r6, r8, r9} work has been carried out
recently on inhomogeneous and/or non-autonomous NLSE in an external potential
due to its physical and experimental relevance. Such systems appear in the study
of Bose-Einstein condensation, soliton laser, ultra-fast soliton switches and
logic gates\cite{ag}. The time dependence of different coefficients may arise due to
time-dependent external forces, whereas inhomogeneity may be introduced through
optical control of Feshback resonances\cite{r7}. One may use the method of
similarity transformation\cite{1jbb,jbb} to find exact solutions of such
inhomogeneous and non-autonomous NLSE and there are many such exactly
solvable systems.

A new integrable non-local NLSE was introduced in Ref. \cite{abm} for
which exact solutions were obtained through inverse scattering method.
In contrast to the standard formulation of NLSE, the Schr$\ddot{o}$dinger
field and its parity(${\cal{P}}$)-transformed complex conjugate are treated
as two independent fields. The self-induced potential in the corresponding
stationary problem is non-hermitian, but, ${\cal{PT}}$ symmetric. It was shown
later\cite{aks} that this non-local NLSE admits both dark and bright solitons
for the case of attractive interaction. Several periodic soliton solutions of
this equation have been obtained analytically\cite{akas}. A two-component
generalization of the non-local NLSE is considered in Ref. \cite{akas}, while
non-local NLSE on one dimensional lattice is introduced in
Refs. \cite{aks,akas}. 

The purpose of this paper is to introduce and study an inhomogeneous and
non-autonomous version
of the integrable non-local NLSE of Ref. \cite{abm}. In particular, we consider
a class of non-local NLSE in an external potential with space-time modulated
coefficient of the nonlinear interaction term as well as confining
and/or
loss-gain terms. We find exact soliton solutions for this generalized class of
non-local NLSE by using a similarity transformation. We find that only those
transformations are allowed for which the transformed spatial coordinate is
odd under the parity transformation of the original one. This is in contrast
to the findings of similar studies for the local NLSE for which no such
restriction is necessary.  Although such a condition puts restriction on the
possible types of external potentials, loss/gain terms, space-time modulated
co-efficients etc., the choices are still infinitely many including most of the
physically interesting cases. We consider a few examples with the explicit
expressions for the external potential and space-time modulated coefficient
of the non-linear interaction term. It is worth mentioning here that
integrability of non-local NLSE with spatiotemporally varying coefficients of
the dispersion as well as the non-linear term has been considered recently by
using Lax-pair formulation \cite{mr}. However, the integrability condition in
Ref. \cite{mr} restricts the spatial dependence of the co-efficients to a
specific form and can not reproduce the class of non-local NLSE  considered
in this paper.

We introduce a $d+1$ dimensional homogeneous and autonomous non-local
NLSE without any external potential and study its Schr$\ddot{o}$dinger
invariance. The system is invariant under all the symmetry transformations
associated with the $d+1$ dimensional Schrodinger group. We find that the
formal expressions for the corresponding conserved Noether charges are
non-hermitian. However, the conserved charges associated with the
time-translation, dilatation and special conformal transformation are shown
to be real-valued only. On the other hand, the total momentum as well as the
boost are complex in any spatial dimensions. Consequently, the angular momentum
turns out to be real in odd spatial dimensions and is complex in even spatial
dimensions. The conserved charges are shown to satisfy the $d+1$ dimensional
Schr$\ddot{o}$inger algebra.

Finally, we consider an inhomogeneous and non-autonomous version of this higher
dimensional non-local NLSE. We introduce different moments and study their
dynamics. Although the formal expressions for these moments are non-hermitian,
they are shown to be real-valued. We find an exact description of the
time-evolution of the `pseudo-width' of the wave-packet for the special case
when the system admits a $O(2,1)$ conformal symmetry.

\section{Exact solution of non-local NLSE}

A new integrable non-local NLSE in 1+1 dimensions has been introduced in
\cite{abm}:
\be
i\psi_{t}({x}, t)=-\frac{1}{2}\psi_{xx}({x},
t)+ G \ \psi^{*}(-x,t)\psi(x,t)\psi(x,t), \ G \in R.
\label{aNLSE1}
\ee
The self-induced  potential in the corresponding stationary problem has the form,
$V(x)=\psi^{*}(-x) \\ \psi(x)$, which is $\cal{PT}$ symmetric, i.e. 
$V^*(-x)=V(x)$. The equation is non-local in the sense that the value
of the potential $V(x)$ at $x$ requires the information on $\psi$ at $x$
as well as at $-x$. It has been shown in Ref. \cite{abm} that this equation
possess a Lax pair and an infinite number of conserved quantity and
therefore, is integrable. In contrast to the usual local NLSE , eq.
(\ref{aNLSE1}) admits dark as well as bright soliton solutions for
$g <0$\cite{aks}. Several periodic soliton solutions of this equation have also
been found\cite{akas}. It is interesting to note that eq. (\ref{aNLSE1})
do admit solution with special shift in coordinate $x$, but, not with
arbitrary shift \cite{aks,akas}.
 
In this section, we investigate the possible exact solutions of the
following non-autonomous NLSE:
\bea
i\psi_{t}&=&-\frac{1}{2}\psi_{xx}+[V(x,t)+iW(x,t)]\psi+
g(x,t){\psi^{*}}^{p}(-x,t){\psi}^{p}(x,t)\psi(x,t), \ p \in N,
\label{cNLSE}
\eea
\noindent where  $g(x,t)$ is the space-time dependent strength of the
nonlinear interaction. It may be noted that the non-linear
interaction term is non-local as well as ${\cal{PT}}$ symmetric.
The external potential $v(x,t)=V(x,t)+i W(x,t)$ is chosen to be complex
with $V(x,t)$ and $W(x,t)$ being the real and imaginary parts,
respectively. The effect of $V(x,t)$ is to confine the particle, whereas
$W(x,t)$ is considered as gain/loss coefficient. The external potential
$v(x,t)$ becomes ${\cal{PT}}$ symmetric for $V(x,t)=V(-x,-t)$ and
$W(x,t)=-W(-x,-t)$.

The above equation reduces to a homogeneous non-local NLSE.,
\be
i\psi_{t}=-\frac{1}{2}\psi_{xx}+G{\psi^{*}}^{p}(-x,t){\psi}^{p}(x,t)\psi(x,t),
\label{NLSE1}
\ee
\noindent for  $V(x,t)=W(x,t)=0$, $g(x,t)=G$. A further choice of $p=1$
reproduces the non-local NLSE in eq. (\ref{aNLSE1}), which is exactly solvable.
We find an exact solution of eq. (\ref{NLSE1}) for arbitrary $p$,
\bea
\psi(x, t)&=&\Phi_0\ e^{i \frac{A^2}{2p^2} t} \ sech^{\frac{1}{p}}(A x),
\label{arbp}
\eea
where
\be
G=-\frac{A^2(1+p)}{2p^2\Phi^{2p}_0},
\label{Gvalue}
\ee
\noindent is necessarily negative. It may be noted that unlike the soliton
solutions of the corresponding local NLSE, an arbitrary constant shift of the
transverse co-ordinate in $\psi(x)$ does not produce an exact solution of
(\ref{NLSE1}). The known bright soliton solutions of non-local NLSE with cubic
nonlinearity may be reproduced by putting $p=1$ in eqs. (\ref{arbp}) and
(\ref{Gvalue}).

\par We use the similarity transformation \cite{jbb}
\begin{eqnarray}
\psi(x,t)&=&\rho(x,t)e^{i\phi(x,t)} \Phi(X),\  \  \  \ X \equiv X(x,t),
\label{Tr}
\end{eqnarray} 
to map eq. (\ref{cNLSE}) to the following equation:
\begin{eqnarray}
\mu \Phi(X)&=&-\frac{1}{2} \Phi_{XX}(X)+G {\Phi^{*}}^{p}(-X){\Phi}^{p}(X){\Phi}(X).
\label{snlNLSE}
\end{eqnarray}
Consequently,  the known exact solution of eq. (\ref{NLSE1}) may be used to construct a large class of exactly
solvable non-autonomous non-local  NLSE  of the type of eq. (\ref{cNLSE})..
We  find  that eq. (\ref{cNLSE}) reduces to the stationary non-local NLSE (\ref{snlNLSE}) only when
$X(x, t)$ is an odd function of $x$, i.e.,
\be
 X(-x, t)=-X(x, t), 
\label{TrX}
\ee
and  the following additional consistency conditions hold simultaneously:
\bea
2\rho \rho_t +(\rho^2\phi_x)_x&=&2\rho^2 W(x,t)
\label{e1}\\
(\rho^2 X_x)_x&=&0
\label{e2}\\
X_t +\phi_x X_x &=&0
\label{e3}\\
V(x,t)&=&\frac{\rho_{xx}}{2\rho}-\phi_t-\frac{\phi^2_x}{2}-\mu X^2_x
\label{e4}\\
g(x,t)&=&\frac{G}{\rho^p(-x,t)\rho^p(x,t)e^{ip(\phi(x,t)-\phi(-x,t))}}X^2_x
\label{e5}
\eea
The above conditions are obtained by  exploiting the facts that $\psi$ and $\Phi$ satisfy equations (\ref{cNLSE})
and (\ref{snlNLSE}), respectively  and are related by  transformation in eq.  (\ref{Tr}). It may be noted that
the oddness of $X(x, t)$ in $x$, as in eq. (\ref{TrX}), is not necessary for the similarity transformation from
local NLSE to its inhomogeneous counter part. The condition (\ref{TrX}) solely arises due to the non-local
nature of the nonlinear interaction and forbids any purely time-dependent shift in the choice of $X$ in terms of
$x$ and $t$. This is consistent with the fact that the solutions of the non-local NLSE are not invariant under
any shift of the  transverse co-ordinate $x$\cite{aks}. All the consistency conditions in eqs. (\ref{e1} -
\ref{e5}), except for the expression of $g(x, t)$, are identical with the corresponding expressions\cite{jbb}
obtained for the mapping of local NLSE to its inhomogeneous counterpart. Further, it is evident that $g(x,t)$
becomes a complex function if $\phi(x,t)$ is not an even function, of $x$, while the consistency conditions stated
above are based on the assumption of real $g(x,t)$. This apparent contradiction is removed by the use of
eq. (\ref{TrX}), which reduces $g(x,t)$ to be real. To this end, we  solve eqs. (\ref{e2}) and (\ref{e3}) to
obtain $\rho$ and $\phi$:
\begin{eqnarray}
\rho(x,t) &=& \sqrt{\frac{\delta(t)}{ X_x}},
\ \ \phi(x,t)= - \int dx \frac{X_t}{X_x} + \phi_0(t), 
\label{e}
\end{eqnarray}
where $\delta(t)$ and $\phi_0(t)$ are two time-dependent integration constants. It immediately follows that both
$\rho$ and $\phi(x,t)$ are even in $x$, which allows to re-write $g(x,t)$ in eq. (\ref{e5}) as,
\begin{eqnarray}
g(x,t)&=&\frac{G\delta^2(t)}{\rho^{2(p+2)}}.
\label{e51}
\end{eqnarray}
A choice of $X$ will determine $\rho$ and $\phi$ through eq. (\ref{e}) up to  two time dependent
integration constants which may be fixed by using appropriate conditions on $V(x,t)$.  The expressions of
$X$, $\rho$ and $\phi$ may be used to determine $W(x,t)$, $V(x,t)$ and $g(x,t)$ from equations (\ref{e1}),
(\ref{e4}), and (\ref{e51}) respectively.

\subsection{Inhomogeneous autonomous non-local NLSE}

Consider a  spacial class of similarity transformation by considering,
 \be
\rho(x, t)\equiv \rho(x), \phi(x,t) \equiv -E t, \ X \equiv X(x), 
\ee
in eq. (\ref{Tr}). In this case,  eq. (\ref{e3}) is satisfied automatically
and the consistency condition of eq (\ref{e1}) determines  $W(x, t)=0$, 
which implies that no gain/loss term can be generated under this similarity
transformation. From eqs.  (\ref{e2}), (\ref{e4}) and (\ref{e51}) $X(x)$,
$g(x)$ and $V(x)$ can be determined as,
\bea
X(x)=   \int^{x}_o    \frac{ds}{\rho^{2}(s)}
\label{x}\\
g(x)=\frac{ G }{\rho^{2(p+2)}}
\label{g}\\
V(x)=\frac{\rho_{xx}}{2\rho}+E- \frac{\mu}{\rho^{4}}
\label{v}
\eea
Eq. (\ref{x}) implies that $\rho$ must have a definite parity as $X$ is an
odd function of $x$. It immediately follows from eqs. (\ref{g}) and
(\ref{v}) that both $g(x)$ and $V(x)$ must be an even function of $x$. In
particular, 
\be
\rho(-x)=\pm \rho(x), \ g(-x)=g(x), \  V(-x)=V(x).
\label{rvg}
\ee
The reality of $\rho(x)$, $g(x)$ and $V(x)$ ensures that these
functions are also ${\cal{PT}}$-symmetric. It may be noted that for the
similarity transformation of the local  NLSE to its inhomogeneous
counterpart\cite{1jbb}, no conditions as in eqs. (\ref{TrX}) and
(\ref{rvg}) are necessary. Thus, {\it we have the important result that the
similarity transformation technique\cite{1jbb} is applicable to the
non-local NLSE, only when both the confining potential $V(x)$ and the
space-modulated nonlinear interaction term $g(x)$ are even in $x$}.

\par The expressions for $X(x)$ and $g(x)$ can be obtained, once an
 explicit form of $\rho(x)$ is known. We use eq. (\ref{v}) to find
 $\rho(x)$ for a given $V(x)$. We re-write eq. (\ref{v}) as,
\be
\frac{1}{2} \rho_{xx}+ \left [ E-V(x) \right ] \rho = \frac{\mu}{\rho^3}
\label{ep}
\ee
which is the Ermakov-Pinney equation \cite{1jbb}. The solution of this
equation may be written as,
\be
\rho=\left[a \phi^{2}_1(x) + 2 b \phi_1(x) \phi_2(x) +
c \phi_2^2(x) \right ]^{\frac{1}{2}},
\label{so}
\ee
where a, b, c  are constants and $\phi_1(x)$, $\phi_2(x)$ are the two
linearly independent solutions of the equation,
\be
-\frac{1}{2} \phi_{xx}+V(x)\phi(x)=E\phi(x)
\label{s}
\ee
The constant $\mu$ is determined as,
$\mu= (ac-b^2) \left [ \phi_1^{\prime}(x) \phi_2(x) -\phi_1(x)
\phi_2^{\prime}(x) \right ]^2 $.
The confining potential $V(x)$ has even parity. Thus, $\phi_{1,2}(x)$ can
always be chosen to be either even or odd. The requirement of a definite
parity for $\rho(x)$ can always be ensured by suitably choosing the constants
$a, b, c$ for a given set of linearly independent solutions $\phi_1$ and
$\phi_2$.

\subsubsection{Examples}
We consider a few specific examples.
\begin{enumerate}
\item {\bf Vanishing External Potential}\\

The first example deals with the case of no external potential, i. e.,
${ V(x)=0}$. There are two cases depending on whether $E >0$ or $E <0$, which
are treated separately.
For $E >0$, eqs. (\ref{x}-\ref{s}) can be solved consistently leading to the
following expressions for the function $\rho(x)$ and the space-modulated
co-efficient $g(x)$:
\be
\rho(x) = \left [ 1 + \alpha \cos (\omega x) \right ]^{\frac{1}{2}}, \
g(x,t) =  G \left [ 1 + \alpha \cos (\omega x) \right ]^{-(p+2)},
\label{rhog}
\ee
where $\omega= 2 \sqrt{2 {\mid E \mid}}$ and $\mu=(1-\alpha^2) E$. The
transformed co-ordinate $X(x)$ is determined as,
\bea
X_+(x) & = & \frac{2}{\omega \sqrt{1-\alpha^2}} tan^{-1} \left [
\sqrt{\frac{1-\alpha}{1+\alpha}} \tan(\frac{\omega x}{2} )\right ] \ for \
{\mid \alpha \mid} < 1,\nonumber \\
X_-(x)&=& \frac{1}{\omega \sqrt{\alpha^2-1}} ln \left [\frac{ 
tan(\frac{\omega x}{2}) + \sqrt{\frac{\alpha+1}{\alpha-1}}}
{ tan(\frac{\omega x}{2}) - \sqrt{\frac{\alpha+1}{\alpha-1}}}
 \right ] \ for \ {\mid \alpha \mid} > 1,
\eea
where the subscripts refer to the fact that $\mu$ is positive for the solution
$X_+(x)$, whereas it is negative for $X_-(x)$.  A solution of eq. (\ref{cNLSE})
for $G <0$, $V=W=0$ and $g(x,t)$ given by eq. (\ref{rhog}) reads,
\be
\psi(x,t) = e^{-i E t} \left ( \frac{E(\alpha^2-1)(p+1)}{{\mid G \mid}}
\right )^{\frac{1}{2p}}
\left [ 1 + \alpha \cos (\omega x) \right ]^{\frac{1}{2}} \
sech^{\frac{1}{p}} \left ( p \sqrt{2 E (\alpha^2-1)} X_-(x) \right )
\ee
For $p=1$, under the same conditions stated above, eq. (\ref{cNLSE}) also
admits the solution:
\be
\psi(x,t) = e^{-i E t} \left ( \frac{E(1-\alpha^2)}{\mid G \mid} 
 \left [ 1 + \alpha \cos (\omega x) \right ] \right )^{\frac{1}{2}} \
tanh \left ( \sqrt{ (1-\alpha^2) E } X_+(x) \right )
\label{dark}
\ee
It may be noted that Eq. (\ref{dark}) is also a solution of the corresponding
local NLSE, but, for $ G >0$.
 
For $E < 0$, eqs. (\ref{x}-\ref{s}) can be solved consistently with the
following expressions for the function $\rho(x)$, the space-modulated
co-efficient $g(x)$ and the transformed co-ordinate $X(x)$:
\be
\rho(x)= \cos h^{\frac{1}{2}} (\omega x), \ \
g(x) =  G cosh^{-(p+2)}(\omega x), \
X(x)= - \frac{1}{\omega} cos^{-1} (tanh (\omega x))
\ee
where $\mu$ is determined as $\mu= 2 {\mid E \mid}$ which is positive-definite.
Unlike $x$ which is defined on the whole line,  $X$ is bounded within the
range $ 0 \leq X \leq \frac{\pi}{\omega}$ and any solution of eq.
(\ref{snlNLSE}) must vanish at the end points. There are many exact
periodic solutions\cite{akas} of eq. (\ref{snlNLSE}) for $p=1$ in terms of
Jacobi elliptic functions. The type-$V$ and type-$VIII$  solutions of Ref.
\cite{akas} are of particular interests to the present problem. In particular,
\be
\psi_V=e^{-i E t} \left ( \frac{2 m \mu}{{\mid G \mid} (1+m)} cosh(\omega x)
\right )^{\frac{1}{2}} sn(\sqrt{\frac{2\mu}{1+m}} X, m) 
\ee
is an exact solution of Eq. (\ref{cNLSE}) with $p=1$ and $G < 0$, where
$ \frac{1}{2} < \mu \leq 1$. The value of $m$ within the range $ 0 < m \leq 1$ is
determined from the condition
\be
4 n K(m) \sqrt{a_m} =\pi, \ \
K(m) \equiv \int_0^{\frac{\pi}{2}} (1-m sin^2 \theta)^{-\frac{1}{2}} d\theta,
\label{condi}
\ee where $n$ is any positive integer and $a_m=m+1$.
The above equation determining the allowed values of $m$ arises from the
condition that $sn(\sqrt{\frac{2\mu \pi}{(1+m)\omega}}, m)=0$ and for every
$n$ it has a unique solution\cite{1jbb}. The boundary condition at $X=0$ is
automatically satisfied by the elliptic function
$sn(\sqrt{\frac{2\mu}{1+m}} X, m)$. A second solution of Eq. (\ref{cNLSE})
with $p=1$ and $G > 0$ is,
\be
\psi_{VIII}=e^{-i E t} \left ( \frac{2 m \mu(1-m)}{{\mid G \mid} (2m-1)}
cosh(\omega x) \right )^{\frac{1}{2}}
\frac{ sn(\sqrt{\frac{2\mu}{1-2m}} X, m)}{ dn(\sqrt{\frac{2\mu}{1-2m}} X, m)} 
\ee
where the values of $m$ within the range $0 < m <\frac{1}{2}$ is again
determined from the equation (\ref{condi}) with $a_m=1-2m$. Both $\psi_V$
and $\psi_{VIII}$ describe bound states of multi-soliton states. 
The inhomogeneous local NLSE corresponding to eq.  (\ref{cNLSE}) also admits
these novel states\cite{1jbb}, but for $G<0$.

\item {\bf Harmonic Confinement}\\

We choose  $V(x)=\frac{1}{2} x^2$ and $E=0$ for which (\ref{x}-\ref{s}) can be
solved consistently with the following solutions:
\be
\rho(x)=e^{\frac{x^2}{2}}, \
g(x)=  G e^{-(p+2) x^2}, \
X(x)= \frac{\sqrt{\pi}}{2} erf x.
\ee
Note that $\mu=0$ and $-\frac{\sqrt{\pi}}{2} \leq X \leq \frac{\sqrt{\pi}}{2}$.
We choose $p=1$ for which solutions of type-II and type-VIII of
Ref. \cite{akas} with $m=\frac{1}{2}$ are relevant for the present discussion.
In particular,
\bea 
\psi_{II}^n & =&  \frac{2 n K(\frac{1}{2})}{\sqrt{2\pi {\mid G \mid}}}
e^{-i Et} e^{\frac{x^2}{2}}
cn(\theta_n, \frac{1}{2}), n=1, 3, \dots \nonumber \\
\psi_{VIII}^n & = & \frac{ n K(\frac{1}{2})}{{\sqrt{\pi \mid G \mid}}}
e^{-i Et} e^{\frac{x^2}{2}} \frac{sn(\theta_n, \frac{1}{2})}{dn(\theta_n,
\frac{1}{2})}, \ n=2, 4, \dots
\label{rbp}
\eea
are solutions of eq, (\ref{cNLSE}) for $G <0$ and $G >0$, respectively, where
$\theta_n$ is defined as,
\be
\theta_n(x)= n K(\frac{1}{2}) erf x, \ \ n=1, 2, \dots.
\ee 
It may be recalled that both $\psi_{II}$ and $\psi_{VIII}$ are solutions of
the corresponding local NLSE for $G <0$\cite{1jbb}. The difference between the
local and the non-local cases arises due to the fact that $cn(X)$ and $dn(x)$
are even functions of $X$, while $sn(X)$ is and odd function of its argument.
Both $\psi_{II}$ and $\psi_{VIII}$ are localized in space and each ot them has
$n-1$ zeroes for a fixed $n$\cite{1jbb}.

\item {\bf Reflection-less Potential}

We choose $E=0$ and the potential
\be
V(x)=\frac{1}{2} A^2 - \frac{1}{2}A(A+1) sech^2x, \ \ A \in N,
\ee
for which 
\bea
\rho(x) & = & \left ( cosh \ x \right )^{A}, \
g(x) =  G  \left ( sech \ x \right )^{2A(p+2)} \nonumber \\
X(x) & = & \sum_{k=0}^{A-1} \frac{(-1)^k}{2k+1} \ {^{A-1}C_k} \
\left ( tanh \ x \right )^{2k+1} 
\eea
are consistent with equations (\ref{x}-\ref{s}) and $\mu$ is determined as
$\mu=0$. The range of $X$ is given by
$ - L \leq X \leq L, \ L=\sum_{k=0}^{A-1} \frac{(-1)^k}{2k+1} \ {^{A-1}C_k}$.
We choose $p=1$ for which, 
\bea 
\psi_{II}^n & =&  \frac{ n K(\frac{1}{2})}{L \sqrt{ {2\mid G \mid}}}
e^{-i Et}  \left ( cosh \ x \right )^{A}
cn(\chi_n, \frac{1}{2}), n=1, 3, \dots \nonumber \\
\psi_{VIII}^n & = & \frac{ n K(\frac{1}{2})}{L {2\sqrt{ \mid G \mid}}}
e^{-i Et} \left ( cosh \ x \right )^{A}
\frac{sn(\chi_n, \frac{1}{2})}{dn(\chi_n,
\frac{1}{2})}, \ n=2, 4, \dots
\eea
are solutions of eq, (\ref{cNLSE}) for $G <0$ and $G >0$, respectively, where
$\chi_n$ is defined as,
\be
\chi_n(x)= \frac{\sqrt{\pi}n K(\frac{1}{2})}{2 L} X(x), \
n=1, 2, \dots
\ee 
Both the solutions are localized in space and each of them has $n-1$ zeroes
for fixed $n$.
\end{enumerate}
\subsection{ Non-autonomous non-local NLSE}

The condition (\ref{TrX}) can be implemented in several ways. 
We discuss two different classes of $X(x,t)$ depending on its
separability or non-separability in terms of its arguments $x$ and $t$.
It turns out that for the non-separable case gain/loss co-efficient is
essentially zero, while it may be chosen to be non-zero for the separable
case.\\ 
\subsubsection{Non-separable $X(x,t)$}
One may choose the following ansatz,
\be
X(t,x)=F(\xi), \ \xi(t,x) \equiv \gamma(t) x, \ F(-\xi) = -F(\xi),
\label{oddcondi}
\ee
where $\gamma(t)$ is an arbitrary function of $t$. Note that unlike in the
case of local NLSE\cite{jbb}, a purely time-dependent term can not be
added to the ansatz for $X(x,t)$ due to the condition (\ref{TrX}). Further,
the consistency of the eqs. (\ref{e4}-\ref{e}) fixes $W(x,t)=0$. Thus, the
above ansatz is not suitable for systems with loss/gain term. We obtain the
following expressions:
\bea
\phi(x,t) & = & -\frac{\gamma_t}{2 \gamma} x^2 + \phi_0(t), \ \
\rho(x,t) = \sqrt{\frac{\gamma}{F^{\prime}(\xi)}},\ \
g(x,t) = { G \gamma^{2-p}} \left ( F^{\prime}(\xi) \right )^{p+2}
,\nonumber \\
V(x,t) & = & \frac{\gamma^2}{8 (F^{\prime}(\xi))^2} \left [
3 \left \{ F^{\prime \prime}(\xi) \right \}^2 - 2 F^{\prime \prime}(\xi)
 F^{\prime \prime \prime}(\xi) - 8 \mu \left \{ F^{\prime}(\xi)
\right \}^4 \right ] + \frac{1}{2} \omega(t) x^2 - \phi_{0t}
\eea
where we have assumed $\delta=\gamma^2$ and for a given $\omega(t)$,
$u=\gamma^{-1}$  is determined from the equation:
\be
u_{tt} + \omega(t) u=0.
\label{mathieu}
\ee
\noindent The above ansatz leads to harmonic confinement irrespective of the
choice of $F(\xi)$. It may happen that the first term in $V(x,t)$ contains a
term proportional to $\xi^2$ for specific choices of $F(\xi)$ for which eq.
(\ref{mathieu}) gets transformed into the Ermakov-Pinney equation with \cite{jbb}.

The example considered in Ref. \cite{jbb} for the case of corresponding local
NLSE with $p=1$ is that of exponentially localized non-linearity with a combination of
harmonic and dipole traps. The motivation behind such a choice is the 
experimental scenario related to Bose-Einstein condensation. It may be noted
that $F(\xi)= \int e^{-\xi^2} d \xi$ is an odd function of its arguments and
satisfies the conditions (\ref{oddcondi}). Thus, the results of
Ref. \cite{1jbb}) are equally valid for the non-local NLSE with $p=1$ also, except for the
following differences:\\
(i) The discussions in Ref. \cite{1jbb} for local NLSE are for attractive
interaction ($G=-1$), whereas the same results are valid for the non-local
NLSE under consideration for repulsive interaction($G=1$) only.\\
(ii) The non-local NLSE admits resonant soliton, breathing soliton and
quasi-periodic solutions. However, moving solitons are not allowed for
the non-local NLSE due to the condition (\ref{TrX}) which forbids the addition
of a purely time-dependent term to the ansatz for $X(x,t)$.

\subsubsection{Separable $X(x,t)$}
We choose an expression for $X(x,t)$ which is separable in terms of its
arguments and the spatial part is an odd function of $x$:
\be
X(x,t) \equiv \alpha(t)f(x), \ \ f(-x)= - f(x).
\ee
With this choice of $X$, eqs. (\ref{e}-\ref{e4}) take the following form in
terms of $\alpha(t)$ and $f(x)$:
\begin{eqnarray}
\rho(x,t) &=&\sqrt{\frac{\delta(t)}{\alpha(t)f'(x)}}
\label{exrho}, \ \
\phi(x,t) = -\frac{\alpha_t}{\alpha(t)} \int dx \frac{f(x)}{f'(x)}, \
g(x,t) = \frac{ G \alpha^{p+2}}{\delta^{p}} {(f^{\prime})}^{p+2}
\label{exphi}\\
W(x, t)&=&\frac{1}{2\alpha(t)\delta(t)}(\delta_t\alpha-2 \alpha_t\delta)
+ \frac{\alpha_t}{\alpha}(\frac{f'' f}{f'^2}),
\label{exW}\\
V(x, t)&=&-(\frac{2f'''f'-3f''^2}{8f'^2})+\frac{\alpha_{tt}\alpha-
\alpha^2_t}{\alpha^2}\int \frac{f(x)}{f'(x)}dx-
\frac{\alpha^2_tf^2}{2\alpha^2f'^2}-\mu\alpha^2f'^2,
\label{exV}
\end{eqnarray}
\noindent where $f^{\prime}(x)=\frac{df}{dx}$. We have chosen $\phi_0(t)$
to be zero, since its sole effect is to add a purely time-dependent term
to $V(x,t)$, which can always be removed through a phase rotation.
Following points are in order at this point:\\
\begin{itemize}
\item {\bf ${\cal{PT}}$-symmetry}: It may be noted that $W$ is odd and $V$ is
even under ${\cal PT}$,
whenever both $\alpha(t)$ and $\delta(t)$ have definite parity. Thus, this
is also the condition for the external potential $v(x,t)$ to be ${\cal PT}$
symmetric. The space-time modulated nonlinear interaction $g(x,t)$ becomes
${\cal PT}$ symmetric, when additional conditions are imposed. In
particular, it becomes ${\cal PT}$  symmetric when  both $\delta(t)$ and
$\alpha(t)$ have the same parity or $p$ is even.

\item {\bf Parameter fixing}: A purely time-dependent part
$W_0(t)=\frac{1}{2\alpha(t)\delta(t)}
(\delta_t\alpha-2 \alpha_t\delta)$ of $W$ can be gauged way from the
equation (\ref{cNLSE})  through a redefinition of $g(x,t)$:
\be
\psi(x,t) \rightarrow \psi(x,t) \ e^{\int^t W_0(t^{\prime}) dt^{\prime}}, \ \
g(x,t) \rightarrow g(x,t)  e^{2 p \int^t W_0(t^{\prime}) dt^{\prime}}.
\ee
Thus, without any loss of generality, we may choose $\delta(t)=\alpha^2(t)$
so that $W_0(t)=0$. The expression for $V$ and $\phi$ remains unchanged
for this particular choice, while $\rho$, $g$ and $W$ read,
\be
\rho(x,t)= \sqrt{\frac{\alpha}{f^{\prime}(x)}}, \
g(x,t) =  G \alpha^{2-p} (f^{\prime})^{p+2}, \
W(x,t)= \frac{\alpha_t}{\alpha}(\frac{f'' f}{f'^2}).
\ee
The system described by eq. (\ref{NLSE1}) has a conformal symmetry for $p=2$
for which $g(x,t)$ becomes independent of time.

\item {\bf Harmonic confinement:} The loss/gain term $W(x,t)$ is purely
time-dependent for $f(x)$ satisfying the following equation:
\bea
f^{\prime \prime} f=f_0 (f^{\prime})^2, f_0 \in R.
\eea
\noindent The odd solution of the above equation with $f_0= {\frac{2n}{2n+1}}$
is $ f(x)=x^{2n+1}, \ n \in N_0$. For the special choice of $n=0$, 
$V(x,t)$ becomes a purely time-dependent harmonic potential:
\be
V(x,t)= \frac{1}{2} \omega(t) x^2 - \mu \alpha^2, \ \
\ee
with $W(x,t)=0$ and $g(x,t)=  G \alpha^{2-p}$. Note that eq. (\ref{mathieu})
can be used to determine $\omega(t)$ for a given $u=\alpha^{-1}$ or the vice-verse.
The system described by eq. (\ref{NLSE1}) has a conformal symmetry for $p=2$
for which $g(x,t)$ becomes space-time independent.\\

A particular choice may be constant $\omega(t)=\omega^2_0$, the general solution of 
eq. (\ref{mathieu}) in this case  yields:
\begin{equation}
\alpha(t)= \left(C_1\cos (\omega_0t)+C_2\sin(\omega_0t)\right)^{-1}
\end{equation}
where $C_1$, $C_2$ are two arbitrary constants. In this case $g(x,t)$, $\phi(x,t)$, $\rho(x,t)$
 and $X(x,t)$ have the following expressions:

\begin{eqnarray}
g(x, t)&=& G \left(C_1\cos (\omega_0t)+C_2\sin(\omega_0t)\right)^{p-2},\ \phi(x,t)
=-\frac{\omega_0\left(C_1\sin(\omega_0t)-C_2\cos(\omega_0t)\right)}
{2\left(C_1\cos (\omega_0t)+C_2\sin(\omega_0t)\right)}x^2\nonumber\\
\rho&=&\left(C_1\cos (\omega_0t)+C_2\sin(\omega_0t)\right)^{-\frac{1}{2}},
\ \ X(x, t)=\left(C_1\cos (\omega_0t)+C_2\sin(\omega_0t)\right)^{-1}x
\label{1aaa}
\end{eqnarray}

We use type $V$ solution of Ref. \cite{akas} to obtain a solution of eq. (\ref{cNLSE})  
with $p=1$ and $G<0$.
\begin{eqnarray}
\psi_{V}&=&\left(\frac{2\mu m}{G(1+m)\left(C_1\cos (\omega_0t)+C_2
\sin(\omega_0t)\right)}\right)^{\frac{1}{2}}e^{-i\frac{\omega_0\left(C_1\sin(\omega_0t)-C_2
\cos(\omega_0t)\right)}{2\left(C_1\cos (\omega_0t)+
C_2\sin(\omega_0t)\right)}x^2} \nonumber\\
&.&sn\left(\sqrt{\frac{2 \mu}{1+m}}X, m\right)\nonumber
\end{eqnarray}
where $\frac{1}{2}<\mu \le 1$ and the value of m within the range $0<m\le 1$ is determined 
from condition given in (\ref{condi}).

Another solution of eq. (\ref{cNLSE}) for $G<0$ and arbitrary $p$ as given by 
eqs. (\ref{arbp}) and (\ref{Tr}) reads,
\be
\psi=\Phi_0\rho e^{i\phi}\ sech^{\frac{1}{p}}(A X)
\ee
where $\rho$, $\phi$ and $X$ are  given by eq. (\ref{1aaa}) and
\be
A=-2\mu p^2 ,\ \ \ \phi_0=\left(\frac{\mu (1+p)}{\mid G\mid}\right)^{\frac{1}{2p}}.
\ee

\item {\bf Non-polynomial external potential:}

A space-time dependent $W(x,t)$ can be produced with non-polynomial
$f(x)$. We choose $f(x)=\sinh x$  for which $g(x,t)$, $W(x,t)$ and $V(x,t)$ 
have the following expressions:
\bea
g(x,t) &=&  G \alpha^{2-p} cosh^{p+2} x,\ \
W(x,t)= \Gamma(t) \ tanh^2 x,\nonumber \\
V(x,t) & = & -\frac{1}{4} + \frac{3}{8} tanh^2 x +
\left ( \frac{d \Gamma}{dt} \right ) ln(cosh x) - \frac{\Gamma^2}{2} tanh^2 x
- \mu \alpha^2 cosh^2 x,
\eea
where $\Gamma(t)=\frac{\alpha_t}{\alpha}$. The function $\psi(x,t)$ reads,
\be
\psi(x,t) = \Phi_0 \sqrt{\alpha} sech^{\frac{1}{2}} x \
exp[ - i \Gamma(t) ln(cosh x) ] \ sech^{\frac{1}{p}}
\left [ A \alpha(t) sinh x \right ], 
\ee
where $A^2=-2 \mu p^2$,   $\Phi_0=\left(\frac{\mu (1+p)}{\mid G\mid}\right)^{\frac{1}{2p}}.$

\end{itemize}

\section{Schr$\ddot{ o}$dinger invariance of non-local NLSE}

A  $d+1$ dimensional generalization of (\ref{aNLSE1}) may be written as
\bea
i\psi_{t}({\bf x}, t)&=&-\frac{1}{2}{\nabla}^{2}\psi({\bf x}, t) +
g \left \{ \psi^*({\cal{P}}{\bf x},t) \psi({\bf x}, t) \right \}^p
\psi({\bf x}, t).
\label{1NLSE2}
\eea
The potential in the corresponding stationary problem has the form,
$V({\bf x})=\left ( \psi^{*}({\cal{P}}{ \bf x}) \psi({\bf x})
\right )^{p}$, which is $\cal{PT}$ symmetric in any spatial dimensions.
It may be recalled that ${\bf x} \rightarrow - {\bf x}$ describes a
rotation in even space dimensions, while  it is parity transformation
in odd spatial dimensions. Thus, $\psi^*(-x,t)$ is replaced with
$\psi^*({\cal{P}}x,t)$ for the higher dimensional generalization of
(\ref{aNLSE1}). The parity transformation in higher
dimensions is not unique and may be parametrized in terms of $d-1$ parameters.
All such parity transformations are related to each other through rotations
in $d$ dimensional space.  One may choose $N$ set of values of these $d-1$
parameters and define the corresponding parity operations as ${\cal{P}}_i,
i=1, 2, \dots N$.  The corresponding ${\cal{PT}}$ symmetric potentials,
\be
\tilde{V}_i({\bf x}) = \left \{ \psi^{*}({\cal{P}}_i{ \bf x})
\psi({\bf x}) \right \}^p, 
\ee
are related to each other through spatial rotation. However, for systems without
rotational invariance, $\tilde{V}_i({\bf x})$'s are to be treated as independent
of each other. For example, if eq. (\ref{1NLSE2}) is considered in an external
potential with space modulated coefficient of the non-linear interaction
term which are not invariant spatial rotation, then each $\tilde{V}_i({\bf x})$
corresponds to different systems.

A Lagrangian formulation of eq. (\ref{1NLSE2}) may be given in terms of the
Lagrangian density,
\bea
{\cal L}&=& i\psi^{*}({\cal{P}}{\bf x}, t)\partial_t\psi(x, t) -
\frac{1}{2}\nabla\psi^{*}({\cal{P}}{\bf x}, t) \cdot
\nabla\psi({\bf x}, t)\nonumber\\
&-&\frac{g}{p+1} \left \{ \psi^{*}({\cal{P}}{\bf x} ,t)\psi({\bf x} ,t)
\right \}^{p+1},
\eea
where $\psi({\bf x} ,t)$ and $\psi^{*}({\cal{P}}{\bf x}, t)$ are treated as two
independent fields. The conjugate momentum associated with
$\psi({\bf x},t)$ is $\Pi_{\psi}({\bf x},t)= i \psi^*({\cal{P}} {\bf x},t)$
and the equal-time Poisson bracket between them leads to the relation:
\be
\left \{ \psi({\bf x},t), \psi^*({\cal{P}} {\bf y}, t) \right \} = - i 
\delta^d({\bf x} -{\bf y}).
\label{poisson}
\ee
It may be recalled that in the Lagrangian formulation of the usual local
NLSE and other field theoretical models involving complex scalar field,
$\psi(x,t)$ and its complex conjugate $\psi^*(x,t)$ are treated as independent
fields. The equal-time Poisson bracket relation between $\psi(x,t)$ and
$\psi^*(x,t)$ in the standard formulation is similar to eq. (\ref{poisson}),
i.e.  $\left \{ \psi({\bf x},t), \psi^*({\bf y}, t) \right \} =
- i \delta^d({\bf x} -{\bf y})$.

The action ${\cal{A}}=\int {\cal L} d^d{\bf x}dt$ is invariant under space-time
translations, spatial rotation, Galilean transformation and a global gauge 
transformation. The action
${\cal{A}}$ is invariant under dilatation and special conformal transformation
for the special case $pd=2$. The symmetries of the action are discussed below:

\begin{description}

\item[1)] {\bf Global $U(1)$ Invariance:}\\
The action ${\cal{A}}$ is invariant under a global $U(1)$ transformation,
$\psi({\bf x},t) \rightarrow \psi'(x, t)=e^{i s}\psi(x, t)$, where $s$ is
a real constant. The corresponding conserved charge is the total number 
$N$,
\be
N=\int  \rho({\bf x},t) d^d{\bf x}, \
\rho({\bf x},t) \equiv \psi^{*}({\cal{P}}{\bf x}, t)\psi({\bf x}, t).
\label{1cq}
\ee
Note that $N$ is neither hermitian nor a semi-positive definite quantity.
Thus, $N$ is identified as quasi-power in the literature\cite{abm}. We now
show that $N$ is real-valued. It is always possible to decompose
$\psi({\bf x}, t)$ as a sum of parity-even and parity-odd terms:
\be
\psi({\bf x}, t)=\psi_e({\bf x}, t)+\psi_o({\bf x}, t),
\label{sa}
\ee
where 
\be
\psi_e({\bf x}, t)=\frac{\psi({\bf x},t)+\psi({\cal P}{\bf x},t)}{2}, \ \
\psi_o({\bf x}, t)=\frac{\psi({\bf x},t)-\psi({\cal P}{\bf x},t)}{2}.
\ee
\noindent With this decomposition of $\psi({\bf x}, t)$, the density
$\rho$ can be expressed in terms of the re-defined field-variables
as sum of a real-valued parity-even term and a parity-odd term which is purely
imaginary. In particular,
\be
\rho({\bf x}, t)=\rho_r({\bf x},t) + \rho_c({\bf x},t)
\label{AB}
\ee
with
\be
\rho_r({\bf x}, t)={\mid \psi_e({\bf x}, t) \mid}^2 -
{\mid \psi_o({\bf x}, t) \mid}^2, \ \
\rho_c({\bf x}, t)=\psi^{*}_e({\bf x}, t)\psi_o({\bf x}, t)-\psi^{*}_o({\bf x}, t)\psi_e({\bf x}, t).
\label{exAB}
\ee
\noindent 
Note the following properties of $\rho_r({\bf x},t)$ and $\rho_c({\bf x},t)$:
\bea
&& \rho_r^{*}({\bf x},t)=\rho_r({\bf x},t),
\ \ {\cal{P}} \rho_r({\bf x},t)=\rho_r({\bf x},t),\nonumber \\
&& \rho_c^{*}({\bf x},t)= -\rho_c({\bf x},t),
\ \ {\cal{P}} \rho_c({\bf x},t)=-\rho_c({\bf x},t), \
\eea
\noindent The density is a complex-valued function. However, the total number
$N$, as defined by eq. (\ref{1cq}), does not receive any contribution from
the parity-odd purely imaginary term $\rho_c({\bf x},t)$ and is real,
$N=\int d^d{\bf x} \rho_r({\bf x},t)$. This result is valid for any spatial
dimensions and we have illustrated it in the appendix-A for one and two
spatial dimensions.  Note that unlike the local NLSE $N$ can take positive as
well as negative values. Thus, a proper interpretation is required for the
total number operator $N$ in the corresponding quantum theory. 
 
The continuity equation for eq. (\ref{1NLSE2}) reads,
\be
\frac{\partial \rho}{\partial t} + {\bf \bigtriangledown} \cdot {\bf J}=0,
\ 
{\bf J}= \frac{i}{2} [\psi({\bf x}, t) {\bf \nabla} \psi^{*}({\cal{P}}
{\bf x}, t)- \psi^{*}({\cal{P}}{\bf x}, t)\nabla \psi({\bf x}, t)],
\ee
\noindent where the current density ${\bf J}$ can be re-written in terms of
the fields $\psi_e({\bf x},t)$ and $\psi_0({\bf x},t)$ as sum of a parity-odd
real term and a parity-even purely imaginary term,
$ {\bf J}={\bf J}_r+ {\bf J}_i$,
with
\bea
{\bf J}_{r} = \frac{i}{2}\left[\psi_e({\bf x},t)\nabla\psi^{*}_e({\bf x},t)-
\psi_o({\bf x},t)\nabla\psi^{*}_o({\bf x},t)
-\psi^{*}_e({\bf x},t)\nabla\psi_e({\bf x},t)+\psi^{*}_o({\bf x},t)
\nabla\psi_o({\bf x},t)\right],\\
{\bf J}_{i} = \frac{i}{2}\left[\psi_o({\bf x},t)\nabla\psi^{*}_e({\bf x},t)-
\psi_e({\bf x},t)\nabla\psi^{*}_o({\bf x},t)
+\psi^{*}_o({\bf x},t)\nabla\psi_e({\bf x},t)-\psi^{*}_e({\bf x},t)\nabla
\psi_o({\bf x},t)\right].
\eea
\noindent The following properties of ${\bf J}_r$ and ${\bf J}_i$ may be noted,
\bea
&&  {\bf J}_{r}({\bf x},t) ^{*}={\bf J}_{r}({\bf x},t),
\ {\cal{P}} {\bf J}_r({\bf x},t)=- {\bf J}_r({\bf x},t),\nonumber \\
&& {\bf J}_{i}^{*}({\bf x},t)=-{\bf J}_{i}({\bf x},t), \
\ {\cal{P}} {\bf J}_i({\bf x},t)= {\bf J}_i({\bf x},t).
\label{useful}
\eea
\noindent which will be useful in showing real-valuedness of some of the
conserved charges and moments to be defined below.
 
\item[2)] {\bf Spatial translation:}\\
The action is invariant under the spatial translation 
${\bf x'}={\bf x}+\delta {\bf x}$ with
\bea
\psi ^{'}({\bf x'}, t)=\psi ({\bf x}, t), \ \
\psi ^{'*}({-\bf x'}, t)=\psi ({-\bf x}, t),
\eea
giving rise to the momentum ${\bf P}= \int {\bf J} d^d{\bf x}$ as the conserved
charge. The exact solutions of eq. (\ref{1NLSE2}) for $d=1$ are not invariant
under an arbitrary shift of the co-ordinate. Thus, these solutions explicitly
break the translational invariance. Defining the centre of mass location as, 
\be
{\bf X}=\frac{1}{N d} \int {\bf x} \rho({\bf x}, t) d^d{\bf x},
\ee
it is easy to verify by using the continuity equation that,
\be
N \frac{d{\bf X}}{dt}= {\bf P},
\label{mo}
\ee
where ${\mid \frac{d{\bf X}}{dt} \mid}$ may be identified as the speed of the
centre of mass.
The total momentum ${\bf P}$ is complex-valued in even spatial dimensions and
is purely imaginary in odd spatial dimensions. This result is presented in
appendix-A for $d=1,2$. Similarly, one can show that the center of mass
${\bf X}$ is purely imaginary in odd spatial dimensions, while it is complex
in even spatial dimensions. Thus, neither the total momentum nor the center of
mass can be considered as physical.

\item[3)] {\bf Time translation:}\\
The invariance of ${\cal{A}}$ under time translation leads to the
conserved quantity,
\bea
{\cal H}&=&\int \left[\frac{1}{2}\nabla\psi^{*}({\cal{P}}{\bf x}, t)
\cdot \nabla \psi({\bf x}, t) +
\frac{g}{p+1} \left \{ {\psi^{*}}({\cal{P}}{\bf x},t) {\psi}({\bf x},t)
\right \}^{p+1} \right]d^d{\bf x},
\label{hami}
\eea
which is identified as the Hamiltonian of the system.
Note that $H$ is not semi-positive definite, since semi-positivity is not
ensured for none of the terms appearing in ${\cal H}$. The Hamiltonian
is also non-hermitian with the standard definition of norm. This should
be contrasted with the Hamiltonian corresponding to the usual local NLSE
for which $H$ is hermitian and for the defocusing case, it is
semi-positive definite. 

We now show that the total Hamiltonian ${\cal{H}}$ is real-valued in spite of
it being non-hermitian. The kinetic energy term in the Hamiltonian density
can be decomposed as a parity-even real term and a parity-odd purely imaginary
term:
\bea
\g\psi^{*}({\cal P}{\bf x},t).\g\psi({\bf x},t)
& = & \left[{\mid \g\psi_{e}({\bf x},t) \mid }^2 - {\mid \g\psi_{o}({\bf x},t)
\mid }^2 \right ]\nonumber \\
& + & \left[\g\psi_{e}^{*}({\bf x},t).\g\psi_{o}({\bf x},t)-
\g\psi_{o}^{*}({\bf x},t).\g\psi_{e}({\bf x},t)\right]
\eea
\noindent The first term is real and even under parity transformation, while
the second term is purely imaginary and odd under parity transformation.
Thus, the second term does not contribute to ${\cal{H}}$ and the contribution
of the kinetic term to ${\cal{H}}$ is real. Similarly, the interaction term in
${\cal{H}}$ can be shown to be real-valued. In particular,
\bea
\frac{g}{p+1} \int d^d{\bf x} \rho^{p+1} & = & 
\frac{g}{p+1} \int d^d{\bf x} \sum_{j=0}^{p+1} {^{p+1}}C_j \rho_c^j
\rho_r^{p+1-j}\nonumber \\
& = & \frac{g}{p+1} \int d^d{\bf x}
\sum_{k=0}^{[\frac{p+1}{2}]} (-1)^{k} \ {^{p+1}}C_{2k} {\mid \rho_c \mid}^{2k}
\rho_r^{p+1-2k}
\eea
\noindent where $[n]$ denotes the integral part of $n$ and
$^nC_r=\frac{n!}{r!(n-r)!}$. It may be recalled that $\rho_c^j$ is
odd under parity transformation for odd $j$, while $\rho_r^{p+1-j}$ is
parity-even term for any $j$. Thus, the summation over odd $j$ terms
does not contribute to the interaction term. The Hamiltonian ${\cal{H}}$
can be re-written as,
\bea
{\cal{H}} & = & \frac{1}{2} \int d^d{\bf x}
\left[ {\mid \g\psi_{e}({\bf x},t) \mid }^2 -
{\mid \g\psi_{o}({\bf x},t) \mid }^2 \right ]\nonumber \\
& + & \frac{g}{p+1} \int d^d{\bf x}
\sum_{k=0}^{[\frac{p+1}{2}]} (-1)^{k} \ {^{p+1}}C_{2k} {\mid \rho_c \mid}^{2k}
\rho_r^{p+1-2k}
\eea 
\noindent which is real-valued and can take positive as well as negative
values.

\item[4)] {\bf Spatial rotation}:\\
The action is invariant under rotation and the corresponding conserved charge
is the angular momentum whose $d(d-1)/2$ components are given by,
\bea
L_{ij}&=&  \int \left(x_i J_j-x_j J_i\right) d^d{\bf x}, \ \
i, j=1, 2, \dots d,
\eea
\noindent where $J_i$ is the $i$-th component of the current density ${\bf J}$.
It may be verified by using eq. (\ref{useful}) that $L_{ij}$'s are
real in odd spatial dimensions, while complex in even spatial dimensions.

\item[5)]{\bf Galilean transformation:}

The action is  invariant under the Galilean transformation. In particular, the
fields $\psi({\bf x}, t)$, $\psi^*({\cal{P}}{\bf x}, t)$ transform under the
Galilean transformation ${\bf x'}={\bf x}-{\bf v}t$ as,
\bea
\psi^{\prime}({\bf x'}, t)&=&e^{-i{\bf v} \cdot ({\bf x^{\prime}} +
\frac{1}{2}{\bf v}t)}\psi ({\bf x}, t),\\
\psi^{\prime *}({\cal{P}}{\bf x'}, t)&=&e^{i{\bf v} \cdot
({\bf x^{\prime}}+\frac{1}{2}{\bf v}t)} \psi^* ({\cal{P}} {\bf x}, t).
\eea
\noindent It may be recalled that the exact solutions of eq. (\ref{1NLSE2}) for
$d=1$ are not invariant under a purely time-dependent shift of the
co-ordinate. Thus, these solutions break the Galilean invariance explicitly.
The conserved charge associated with the Galilean symmetry is boost,
\bea
{\bf B}= t \ {\bf P} -{\bf X},
\eea
\noindent which is complex-valued in even spatial dimensions and purely
imaginary for odd $d$. The conservation of ${\bf B}$ directly follows from
eq. (\ref{mo}).

\item[6)] {\bf Conformal symmetry for $pd=2$:}

Consider the following transformations:
\begin{eqnarray}
{\bf x}\rightarrow  {\bf  x}_{h}&=&\dot\tau^{-\frac{1}{2}}(t){\bf x}, \  \
t \rightarrow \tau=\tau(t)\nonumber\\
\psi({\bf x}, t)\rightarrow \psi_{h}({\bf x}_{h}, \tau)&=
&\dot\tau ^{\frac{d}{4}}exp(-
i\frac{\ddot\tau}{4\dot\tau}x^{2}_{h})\psi({\bf x}, t)\nonumber\\
\psi^{*}( {\cal{P}} {\bf x}, t)\rightarrow \psi_{h}({-\bf x}_{h}, \tau)
&=&\dot
\tau ^{\frac{d}{4}}exp(i\frac{\ddot\tau}{4\dot\tau}x^{2}_{h})
\psi^{*}({\cal{P}} {\bf x}, t),\nonumber
\end{eqnarray}
where
\be
\tau(t)=\frac{\alpha t +\beta}{\gamma t +\delta}, \ 
\alpha\delta-\beta\gamma=1.
\label{ct}
\ee
The particular choices $\tau(t)=t+\beta$, $\tau(t)=\alpha^{2}t$, and
$\tau(t)=\frac{t}{1+\gamma t}$ correspond to time translation, dilation
and special conformal transformation, respectively. The action ${\cal{A}}$ is
invariant under time-translation in arbitrary $d$ and the  corresponding
conserved quantity is given in eq. (\ref{hami}). The action ${\cal{A}}$ is
invariant under dilatation and special conformal transformations for
$pd=2$. This corresponds to a quintic NLSE in $1+1$ dimensions and
cubic NLSE in $2+1$ dimensions. The conserved charges corresponding to
dilatation$(D)$ and special conformation transformation$(K)$ are,
\bea
D&=& t H-I_2\\
K&=&-t^2 H+2 t D+I_1,
\eea
where the moments $I_1$ and $I_2$ are defined as,
\begin{eqnarray}
I_{1}(t)&=&\frac{1}{2}\int d^d{\bf x} \ x^2 \ \rho({\bf x}, t), \ \
I_{2}(t)=\frac{1}{2}\int d^d{\bf x} \ {\bf x} \cdot {\bf J},
\label{i1-i2}
\end{eqnarray}
\noindent where $ x^2= {\bf x} \cdot {\bf x}$.
$I_1$ may be considered as the `pseudo-width' of the wave packet and $I_2$
represents the growth speed of the system.  It may be noted that neither
$I_1$ nor $I_2$ is hermitian and semi-positive definite. However, both
$I_1$ and $I_2$ can be shown to be real-valued. For example, the moment
$I_1$ may be re-written by using eqs. (\ref{AB}) and (\ref{exAB}) as:
\bea
I_1&=& \frac{1}{2} \int d^d{\bf x} \ x^2 \ \rho_r({\bf x},t)
=\frac{1}{2} \int d^d{\bf x} \ x^2 \ \left [{\mid \psi_e({\bf x}, t) \mid}^2 -
{\mid \psi_o({\bf x}, t) \mid}^2 \right ].  
\eea
\noindent The moment $I_1$ can be expressed as difference of two
semi-positive definite moments, $I_1= I_{1e} - I_{1o}$, where
$I_{1e} \equiv \frac{1}{2} \int d^d{\bf x} \ x^2 \ {\mid \psi_e({\bf x}, t)
\mid}^2$ and
$I_{1o} \equiv \frac{1}{2} \int d^d{\bf x} \ x^2 \ {\mid \psi_o({\bf x}, t)
\mid}^2$.
Unlike the case of local NLSE, $I_1$ can be positive as well as 
negative, which restricts the analysis of its dynamics by using the moment
method. The reality of $I_2$ is explained in appendix-A for $d=1,2$. The
dynamics of $I_{1e}$ and $I_{1o}$ are described in appendix-B.
Finally, it is worth mentioning here that both $D$ and $K$ are real,
since $H$, $I_1$, $I_2$ are all real.

Following Refs. \cite{ae,1ae}, the time-development of $I_1(t)$ can be
determined as,
\be
I_1(t) = \left ( \sqrt{I_1(0)} + \frac{\dot{I_1(0)}}{2 \sqrt{I_1(0)}}
t \right)^2 + \frac{Q}{I_1(0)} t^2, \ \ Q \equiv I_1 H - \left ( \frac{1}{2}
\frac{d I_1}{dt} \right )^2
\ee
\noindent where $I_1(0), \dot{I}_1(0)$ are the values of $I_1(t)$ and
$\frac{dI_1}{dt}$ at $t=0$. The  Casimir operator $Q$ of the underlying
$O(2,1)$ group is a constant of motion and can take real values only. The
moment $I_1$ vanishes at a finite real time $t^*$ for $ Q <0$ only, 
\be
t^*= \frac{4 I_1(0)}{Q+\{\dot{I}_1(0)\}^2} \left [ - \frac{\dot{I}_1(0)}{2}
\pm \sqrt{-Q} \right ].
\ee
\noindent Note that $t^*$ can be made positive by appropriately choosing
$I_1(0)$, $\dot{I}_1(0)$ and $H$. Unlike the case of local NLSE, the vanishing
of $I_1$ at a real finite time does not necessarily imply the collapse of the
condensate. The vanishing of $I_1$ rather signifies a transition from positive
$I_1$ to a negative value or the vice verse. It is not clear at this point
whether this transition has any physical significance or not. The vanishing of
$I_1$ at a finite real time can be achieved when any of the following four
conditions are satisfied:
(i) $ I_1(0) > 0, H < 0$, (ii) $I_1(0) >0, H >0, \dot{I}_1(0) \leq - 2 \sqrt{
I_1(0) H}$, (iii) $ I_1(0) < 0, H > 0$, (ii) $I_1(0) < 0, H < 0,
\dot{I}_1(0) \leq - 2 \sqrt{ {\mid I_1(0) H \mid}}$. The first two conditions
are applicable to the local NLSE also. However, the last two conditions are
valid for the non-local NLSE only.

The action is invariant under a duality symmetry. Consider a particular
$\tau(t)$,
\be
\alpha=\delta=0,\ \gamma=-\frac{1}{\beta}, \
\tau= - \frac{\beta^2}{t},
\ee
\noindent which may be thought of as a combined operation of translation in
time by $\beta$, followed by a special conformal transformation and again
a time-translation by $\beta$. The transformation of the spatial co-ordinate
and the fields read:
\bea
{\bf x}\rightarrow  {\bf  x}_{h}&=&\frac{t}{\beta}{\bf x}
=-\frac{\beta}{\tau}{\bf x},\nonumber \\
\psi({\bf x}, t)\rightarrow \psi_{h}({\bf x}_{h}, \tau)&=
&({\frac{\beta}{t}})^{\frac{d}{2}} exp(i\frac{tx^2}{2\beta^2}
)\psi({\bf x}, t),\nonumber \\
\psi^*({\cal{P}}{\bf x}, t)\rightarrow \psi_{h}^*({\cal{P}}{\bf x}_{h}, \tau)&=
&({\frac{\beta}{t}})^{\frac{d}{2}} exp(-i\frac{tx^2}{2\beta^2}
)\psi^*({\cal{P}}{\bf x}, t),
\eea
\noindent which is known as lens transformation \cite{ddff} for the case
of critical local NLSE. The parameter $\beta$ being real, the theory at
$\tau >0$ is mapped to a theory at a time $t < 0$ and the vice-verse with
$\tau=0=t$ separating the two regions. We choose the following convention,
\be
\beta > 0, \ 0 \leq t \leq \infty, \ -\infty \leq \tau \leq 0.
\ee
\noindent Following Ref. \cite{1ae}, we find
that the system admits explosion-implosion duality either for
(a) $H >0, I_1 (0) >0$ or (b) $H < 0, I_1 (0) <0$ such that $Q >0$, i. e., 
\be
{\mid H \mid} \geq \left ( \frac{\dot{I}_1(0)}{2 \sqrt{\mid I_1(0)\mid}}
\right )^2.
\ee
\noindent The pseudo-width explodes in the physical problem and implodes in
the dual problem for both the cases. The physical problem for the first case
describes the growth of $I_1$ from its initial positive value to $\infty$
at $t=\infty$. On the other hand, for the second case, the initial negative
value of $I_1$ in the physical problem decreases to $-\infty$ at $t=\infty$.
The second case described above is not allowed for the local NLSE, since $I_1$
is a semi-positive definite quantity.
\end{description}

The Noether charges satisfy the $d+1$ dimensional Schr$\ddot{o}$dinger
algebra:
\bea
&& \{H, D\}=H,\ \{H, K\}=2 D, \ \{D, K\}=K,\nonumber \\
&& \{{\bf P}, D\}=\frac{1}{2}{\bf P},  \
\{{\bf P}, K\}={\bf B}, \
\{P_i, L_{jk} \}=-\left ( \delta_{ij} P_k-\delta_{ik} P_j\right ),\nonumber\\
&& \{L_{ij}, L_{kl}\}=-\left ( \delta_{ik}L_{jl}-\delta_{il}L_{jk}-
\delta_{jk}L_{il}+\delta_{jl}L_{ik} \right ) \nonumber \\
&& \{H, {\bf B} \} = {\bf P}, \ \{D, {\bf B} \} = \frac{{\bf B}}{2}, \{P_i, B_j \} =- \delta_{ij}N,
\{B_i  ,L_{jk}, \} =-\left ( \delta_{ij} B_k-\delta_{ik} B_j\right ) ,
\eea
\noindent All other Poisson brackets vanish identically. It may be recalled
that all the conserved charges are non-hermitian. Only $H, D, K$ are
real-valued in any dimensions and $L_{ij}$ are real only in odd spatial
dimensions. Further analysis is required to understand the significance
of this algebra in the context of non-local NLSE.

\section{Dynamics of Moments : }

It is hard to find exact solutions of higher dimensional NLSE or its various
generalizations in its generic form. The exact solution may be found only for
particular cases.  The qualitative nature of solutions of such systems may be
described in terms of the dynamics of various moments\cite{ae,1ae,vmp,rcg}.
In particular, the moments satisfy a set of coupled first-order differential
equations with time as the independent variable. However, in general, this is
not a close system of differential equations and involve spatial integrals
involving fields. An exact time-development of some of the moments may be
described analytically for systems with dynamical conformal
symmetry\cite{ae,1ae}. Consequently, important information regarding the
time-development of the field for different initial conditions may be inferred.

Considering the following non-autonomous NLSE in d+1 dimensions:
\bea
i\psi_{t}({\bf x}, t)&=&-\frac{1}{2}{\nabla}^{2}\psi({\bf x}, t)+
V({\bf x} ,t) \psi({\bf x}, t)\nonumber\\
&+&g({\bf x}, t){\psi^{*}}^{p}({\cal{P}}{\bf x} ,t){\psi}^{p}({\bf x},t)
\psi({\bf x}, t).
\label{bNLSE}
\eea
This is a generalization of eq. (\ref{1NLSE2}) where the system is considered in
an external potential and the constant co-efficient of the nonlinear term is allowed
to become space-time dependent.
We define a moment $H$ in addition to the moments $I_1$ and $I_2$ defined in eqs.
(\ref{i1-i2}):
\begin{eqnarray}
H&=& \frac{1}{2}\int \nabla\psi^{*}({\cal{P}}{\bf x}, t) \cdot
\nabla \psi({\bf x}, t) d^d{\bf x}+ \int G(\rho,{\bf x},t) \ d^d{\bf x},
\label{m3}
\end{eqnarray}
where $G(\rho, x, t)=\frac{g(x,t)}{1+p} \rho^{p+1}$. Defining
$g'=g \rho^p$, $\frac{\partial G}{\partial \rho}=g^{\prime}$.
Following the standard technique\cite{vmp,rcg}, it is straight forward to show
that the moments satisfy the following set of equations:

\begin{eqnarray}
\frac{dI_{1}}{dt}&=&2I_{2},\nonumber \\
\frac{dI_{2}}{dt}&=&-\frac{1}{2}\int \rho({\bf x},t) \left({\bf x}. \nabla V\right)d^d{\bf x}
+\tilde{H}-\frac{1}{2}\int\rho({\bf x}, t) ({\bf x} 
\cdot \nabla g^{\prime}) d^d{\bf x} \nonumber \\
\frac{dH}{dt}&=&-\int \nabla V \cdot {\bf J}d^d{\bf x}+ \int \frac{\partial G}{\partial t}d^d{\bf x}
\label{me3}
\end{eqnarray}
where $\tilde{H}$ is the 1st part of eq. (\ref{m3}). If we restrict to the
quadratic potential of the form $V=\frac{1}{2}\omega^2 {\bf x} \cdot {\bf x}$
and $G=g_0 \rho^{1+\frac{2}{d}}$,
eqs. (\ref{me3}) give a close system of equations:
\bea
\frac{dI_{1}}{dt}&=&2I_{2}
\nonumber \\
\frac{dI_{2}}{dt}&=&-\omega^2 I_1
+H
\nonumber \\
\frac{dH}{dt}&=&-2\omega^2 I_2
\label{c1me3}
\eea
It may be noted that the condition $pd=2$ is essential in deriving the above set
of equations which corresponds to conformal symmetry for the system described by $H$.
A decoupled equation for the pseudo-width ${\cal X}=\sqrt{I_1}$ satisfies the
Hill's equation:
\be
\frac{d^2{\cal X}}{dt^2}+\omega^2{\cal X}=\frac{{Q}}{{\cal X}^3} \ 
\label{hil}
\ee
Eq. (\ref{hil}) has the same form of a particle moving in an
inverse-square potential plus a time-dependent harmonic trap. The general
solution of eq. (\ref{hil}) may be written as,
\be
{\cal X}^{2}(t)=u^{2}+\frac{{Q}}{W^2}v^{2}(t), \ \
W(t) \equiv u\dot v-v \dot u,
\ee
where u(t),v(t) are two independent solutions of the following equation, 
\be
\ddot x +\omega (t)x =0, \ \
u(t_{0})={\cal X}(t_{0}),\dot u(t_0)=\dot{\cal X}(t_0),\dot v(t_0)=0,
 v(t_0) \neq 0
\ee
and $W$ is the corresponding Wronskian. We conclude this section with the
following comments:\\
(i) The system admits explosion-implosion duality\cite{1ae} for the special
choice of the time-dependent frequency $\omega(t) = (\frac{\omega_0
\beta}{t})^2, \omega_0 \in R$ and $Q>0$.\\
(ii) The system exhibits parametric instability\cite{vmp} for periodic
$\omega(t)$ with period $T$ when the condition $\delta={\mid u(T) +\dot{v}(T)
\mid} > 2$ is satisfied with the normalization ${\cal{X}}(0)=0,
\dot{\cal{X}}(0)=1, v(0)=1$. The system is stable for $\delta < 2$.

\section{Summary \& Discussions}
We have considered a generalization of the recently introduced integrable
non-local NLSE with self-induced potential that is ${\cal{PT}}$ symmetric in
the corresponding stationary problem and in contrast to the standard
formulation of complex scalar field theory, the Schr$\ddot{o}$dinger field and
its parity-transformed complex conjugate are treated as two
independent fields. We have studied a class of non-local NLSE in an external
potential with space-time modulated coefficient of the nonlinear interaction
term as well as  confining and/or loss-gain terms. We have obtained
exact soliton solutions for the inhomogeneous and/or non-autonomous
non-local NLSE by using similarity transformation and the method is illustrated
with a few specific examples. We have found that only those transformations are
allowed for which the transformed spatial coordinate is odd under the parity
transformation of the original one. This puts some restrictions on the types of
external potentials, loss/gain terms, space-time modulated co-efficients for
which the method is applicable. Nevertheless, the choices are infinitely many
and most of the physically relevant examples are included. It is interesting
to note that all the solutions of the local NLSE are also solutions of the
corresponding non-local NLSE {\it  with identical space-time modulated
co-efficients, external potential, loss/gain terms, non-linear interaction etc.
}. The difference is that the range of the coupling constant of the nonlinear
interaction term for which the solutions exists is different for an odd
solution of local NLSE and the corresponding non-local NLSE. However, the
ranges are identical for an even solution.

We have studied the invariance of the action of a $d+1$ dimensional
generalization of the non-local NLSE under different symmetry
transformations. We have found that the action is invariant under space-time
translation, rotation, global $U(1)$ gauge transformation and under
 Galilean transformation. The system is invariant under
dilatation and special conformal transformations when $pd=2$. It is shown
that $H, D, K$ and ${\bf L}$ are real-valued, although the formal expressions
of these conserved Noether charges are non-hermitian. The conserved momentum
and the total boost are complex-valued in any spatial dimensions. Further,
the conserved charges satisfy the $d+1$ dimensional Schr$\ddot{o}$dinger
algebra. We have also studied the dynamics of different moments with an exact
description of the time-evolution of the ``pseudo-width'' of the wave-packet
for the special case when the action admits a $O(2,1)$ conformal symmetry.

\section{Acknowledgement}
This work is supported by a grant({\bf DST Ref. NO.: SR/S2/HEP-24/2012})
from Science \& Engineering Research Board(SERB), Department of Science
\& Technology(DST), Govt. of India. {\bf DS} acknowledges a research
fellowship from DST under the same project.

\section{Appendix-A: Real-valuedness of some of the non-hermitian Noether
 charges}

Parity is a discrete transformation with the determinant of the transformation
matrix equal to $-1$. Thus, in odd spatial dimensions, a parity transformation
can be realised by flipping the signs of all the coordinates. On the other hand,
the sign of only an odd number of coordinates can be reversed in the case of
even spatial dimensions. Thus, for example, we have ${\cal{P}} \psi(x, t)=
\psi(-x,t)$ in one spatial dimension.  However, in two spatial dimensions,
we have either ${\cal{P}} \psi(x,y, t)=\psi(-x,y,t)$ or
${\cal{P}} \psi(x,y, t)=\psi(x, -y,t)$. We choose the first relation
as our convention for illustrating results related to the real-valuedness of
some of the conserved Noether charges which are non-hermitian.

\begin{description}
\item [(a)]{\underline {N in d=1 dimension}}
\end{description}

We use the following properties of $\rho_r(x,t)$ and $\rho_c(x,t)$:
\bea
&& \rho_r^{*}({x},t)=\rho_r({x},t),
\ \ {\cal{P}} \rho_r({x},t)=\rho_r({x},t),\nonumber \\
&& \rho_c^{*}({ x},t)= -\rho_c({ x},t),
\ \ {\cal{P}} \rho_c({ x},t)=-\rho_c({ x},t), \
\label{ds}
\eea
\noindent which allows to write $N=\int_{-\infty}^{\infty} dx \rho(x,t)
=\int_{-\infty}^{\infty} dx ( \rho_r(x,t) + \rho_c(x,t) )
=\int_{-\infty}^{\infty} dx \rho_r(x,t)$.

\begin{description}
\item [(b)]{\underline {N in d=2 dimensions}}
\end{description}

Similarly in two dimensions we have:

\bea
{\cal P} \rho_r(x,y, t)= \rho_r(-x,y,t)= \rho_r(x,y,t)\nonumber\\
{\cal P}\rho_c(x,y,t)=\rho_c(-x,y,t)=-\rho_c(x,y,t)
\label{aAB1}
\eea

\begin{eqnarray}
N  &=&\int^{\infty}_{-\infty} \left( \rho_r(x,y,t)+\rho_c(x,y,t)\right)dxdy \nonumber\\
&=&\int^{\infty}_{-\infty}\left[\int^{\infty}_{0}\left( \rho_r(x,y,t)-\rho_c(x,y,t)\right)dx+\int^{\infty}_{0}\left( \rho_r(x,y,t)+\rho_c(x,y,t)\right)dx \right]dy \nonumber\\
&=&\int^{\infty}_{-\infty}\int^{\infty}_{0}2 \rho_r(x,y,t)dxdy
\end{eqnarray}

Thus it turns out that N is real.

\begin{description}
\item [(a)]{\underline {{\bf P} in d=1 dimension}}
\end{description}

We shall  use the following relations:
\begin{eqnarray}
J_{r}({\cal P} x,t)=J_{r}(-x,t)=-J_{r}(x,t)\nonumber\\
 J_{i}({\cal P} x,t)=J_{i}(-x,t)=J_{i}(x,t)
\label{aJ}
\end{eqnarray}

to evaluate the integral

\begin{eqnarray}
P&=&\int^{\infty}_{-\infty}\left[J_r(x)+J_i(x)\right]dx\nonumber\\
\end{eqnarray} 

which turns out to be

\begin{eqnarray}
P&=&\int^{\infty}_{0}\left[-J_r(x)+J_i(x)\right]dx+\int^{\infty}_{0}\left[J_r(x)+J_i(x)\right]dx\nonumber\\
&=&2\int^{\infty}_{0}J_i(x)dx
\end{eqnarray}

\begin{description}
\item [(b)]{\underline {{\bf P} in d=2 dimensions}}
\end{description}

We shall use the following relations:

\begin{eqnarray}
{\cal P}J_{rx}( x,y,t)=J_{rx}(-x,y,t)=-J_{rx}(x,y,t) \nonumber\\
 {\cal P}J_{ix}(x,y,t)=J_{ix}(-x,y,t)=+J_{ix}(x,y,t)\nonumber\\
 {\cal P}J_{ry}(x,y,t)=J_{ry}(-x,y,t)=+J_{ry}(x,y,t)\nonumber\\
 {\cal P}J_{iy}(x,y,t)=J_{iy}(-x,y,t)=-J_{iy}(x,y,t)
\label{aJ1}
\end{eqnarray}

\begin{eqnarray}
{\bf P}&=&\int^{\infty}_{-\infty} \{{\bf J}_r(x,y,t)+{\bf J}_i(x,y,t)
\}dxdy \nonumber\\
&=&\hat{{\bf x}}\int^{\infty}_{-\infty} \{ J_{rx}(x,y,t)+ J_{ix}(x,y,t)
\}dxdy
+\hat{{\bf y}}\int^{\infty}_{-\infty} \{( J_{ry}(x,y,t)+ J_{iy}(x,y,t)
\}dxdy \nonumber\\
&=&\hat{{\bf x}}\int^{\infty}_{-\infty}\int^{\infty}_{0}\left\{  J_{rx}(-x,y,t)+ J_{ix}(-x,y,t)+ J_{rx}(x,y,t)+ J_{ix}(x,y,t)\right\}dxdy \nonumber\\
&+&\hat{{\bf y}}\int^{\infty}_{-\infty}\int^{\infty}_{0}\left\{  J_{ry}(-x,y,t)+ J_{iy}(-x,y,t)+ J_{ry}(x,y,t)+ J_{iy}(x,y,t)\right\}dxdy \nonumber\\
&=&\hat{{\bf x}}\int^{\infty}_{-\infty}\int^{\infty}_{0}\left\{  -J_{rx}(x,y,t)+ J_{ix}(x,y,t)+ J_{rx}(x,y,t)+ J_{ix}(x,y,t)\right\}dxdy \nonumber\\
&+&\hat{{\bf y}}\int^{\infty}_{-\infty}\int^{\infty}_{0}\left\{  J_{ry}(x,y,t)- J_{iy}(x,y,t)+ J_{ry}(x,y,t)+ J_{iy}(x,y,t)\right\}dxdy \nonumber\\
&=&2\int^{\infty}_{-\infty}\int^{\infty}_{0}\left[\hat{{\bf x}}  J_{ix}(x,y,t) +\hat{{\bf y}}  J_{ry}(x,y,t)\right]dxdy \nonumber\\
\end{eqnarray}

\begin{description}
\item [(b)]{\underline {$I_{1}$  in d=1 dimension}}
\end{description}

\bea
I_1&=&\frac{1}{2}\int^{\infty}_{-\infty}x^2\rho(x,t)=\frac{1}{2}\int^{\infty}_{-\infty}x^2[\rho_r(x,t)+\rho_c(x,t)]dx\nonumber\\
&=&\frac{1}{2}\int^{\infty}_{0}\left\{x^2\rho_r(-x,t)+x^2\rho_r(x,t)+x^2\rho_c(-x,t)dx+x^2\rho_c(x,t)\right\}dx\nonumber\\
&=&\int^{\infty}_{0}x^2\rho_r(x,t)dx
\eea

where we have used eqs. (\ref{AB}), (\ref{ds}).

\begin{description}
\item [(b)]{\underline { $I_{1}$  in d=2 dimensions}}
\end{description}

\bea
I_1&=&\frac{1}{2}\int^{\infty}_{-\infty}\left[x^2\rho(x,y,t)+y^2\rho(x,y,t)\right]dxdy=\frac{1}{2}\int^{\infty}_{-\infty}\left(x^2+y^2\right)\left(\rho_r+\rho_c\right)dxdy\nonumber\\
&=&\frac{1}{2}\int^{\infty}_{-\infty}\int^{\infty}_{0}\left\{x^2\rho_r(-x,y,t)+x^2\rho_r(x,y,t)+x^2\rho_c(-x,y,t)+x^2\rho_c(x,y,t)\right\}dxdy\nonumber\\
&+&\frac{1}{2}\int^{\infty}_{-\infty}\int^{\infty}_{0}\left\{y^2\rho_r(-x,y,t)+y^2\rho_r(x,y,t)+y^2\rho_c(-x,y,t)+y^2\rho_c(x,y,t)\right\}dxdy\nonumber\\
&=&\int^{\infty}_{-\infty}dy\left[\int^{\infty}_{0}x^2\rho_r(x,y,t)+y^2\rho_r(x,y,t)\right]dx
\eea

where we have used eq. in (\ref{aAB1}).

\begin{description}
\item [(a)]{\underline { $I_{2}$  in d=1 dimensions}}
\end{description}

\bea
I_2&=&\frac{1}{2}\int^{\infty}_{-\infty}dxxJ=\int^{\infty}_{-\infty}dxx\left(J_{r}(x,t)+J_{i}(x,t)\right)\nonumber\\
&=&\frac{1}{2}\int^{\infty}_{0}\left\{(-x)J_r(-x,t)+xJ_r(x,t)+(-x)J_i(-x,t)+xJ_i(x,t)\right\}dx\nonumber\\
&=&\int^{\infty}_{0}dxxJ_r(x,t)\nonumber\\
\eea

where we have used eq. in (\ref{aJ}).

\begin{description}
\item [(b)]{\underline {$ I_{2}$  in d=2 dimensions}}
\end{description}

\bea
I_2&=&\frac{1}{2}\int^{\infty}_{-\infty}dxdy\left(xJ_x+yJ_y\right)\nonumber\\
&=&\frac{1}{2}\int^{\infty}_{-\infty}dxdy\left[x\left\{J_{rx}(x,y,t)+J_{ix}(x,y,t)\right\}+y\left\{J_{rx}(x,y,t)+J_{ix}(x,y,t)\right\}\right]\nonumber\\
&=&\frac{1}{2}\int^{\infty}_{-\infty}dy\int^{\infty}_{0}\left\{-xJ_{rx}(-x,y,t)+xJ_{rx}(x,y,t)-xJ_{ix}(-x,y,t)+xJ_{ix}(x,y,t)\right\}dx\nonumber\\
&+&\frac{1}{2}\int^{\infty}_{-\infty}dy\int^{\infty}_{0}\left\{yJ_{ry}(-x,y,t)+yJ_{ry}(x,y,t)+yJ_{iy}(-x,y,t)+yJ_{iy}(x,y,t)\right\}dx\nonumber\\
&=&\int^{\infty}_{-\infty}dy\left[\int^{\infty}_{0}xJ_{rx}(x,y,t)+\int^{\infty}_{0}yJ_{ry}(x,y,t)\right]\nonumber\\
\eea

where we have used eq. in (\ref{aJ1}).

\section{Appendix-B: Dynamics of $I_{1e}, I_{1o}, I_{2e}, I_{2o}$}

In this appendix we show that the time derivative of $I_1$ and $I_2$ admit
a partial splitting in terms of $\psi_{e}$ and $\psi_{o}$. We present our
results for $d=1$. However, it can be easily generalized to higher dimensions.
The time-development of $I_{1e}$ and $I_{1o}$ are described by the equations,
\bea
\frac{dI_{1e}}{dt}=2I_{2e}-\frac{ig}{2}\int^{\infty}_{-\infty}x^2\rho^2\left
(\psi^{*}_{e}\psi_{o}+\psi_{e}\psi^{*}_{o}\right)dx
\label{c}\\
\frac{dI_{1o}}{dt}=2I_{2o}-\frac{ig}{2}\int^{\infty}_{-\infty}x
^2\rho^2\left(\psi^{*}_{e}\psi_{o}+\psi_{e}\psi^{*}_{o}\right)dx
\label{d}
\eea
\noindent where the moments $I_{2e}$ and $I_{2o}$ are defined as,
\bea
I_{2e}=\frac{1}{2}\int^{\infty}_{-\infty} dx \ x\ \frac{i}{2}\left(\psi_{e}\frac{
\partial\psi^{*}_{e}}{\partial x}-\psi^{*}_{e}\frac{\partial
\psi_{e}}{\partial x}\right)\\
I_{2o}=\frac{1}{2}\int^{\infty}_{-\infty} \ dx \ x\ \frac{i}{2}\left(\psi_{o}
\frac{\partial\psi^{*}_{o}}{\partial x}-\psi^{*}_{o}\frac{\partial\psi_{o}}
{\partial x}\right)
\eea
\noindent If we subtract eq. (\ref{d}) from eq.  (\ref{c}), then left hand side
gives $\frac{dI_1}{dt}$ and the last terms in the right hand sides cancel,
leading to the equation $\frac{dI_1}{dt}=2I_2$.

The equations satisfied by $I_{2e}$ and $I_{2o}$ are,
\bea
\frac{dI_{2e}}{dt}&=&H_{ke}-\frac{g}{4}\int^{\infty}_{-\infty}x\frac{\partial
\rho^2}{\partial x}
\left(2\psi^{*}_{e}\psi_{e}+\psi_{o}\psi^{*}_{e}-\psi^{*}_{o}\psi_{e}\right)dx
\nonumber\\
&+&\frac{g}{4}\int^{\infty}_{-\infty}x\rho^2\left(\psi_{o}\frac{\partial
\psi^{*}_{e}}{\partial x}-
\psi^{*}_{o}\frac{\partial \psi_{e}}{\partial x}+\psi_{e}\frac{\partial
\psi^{*}_{o}}{\partial x}-
\psi^{*}_{e}\frac{\partial \psi_{o}}{\partial x}\right)dx
\label{a}\\
\frac{dI_{2o}}{dt}&=&H_{ko}-\frac{g}{4}\int^{\infty}_{-\infty}x\frac{\partial
\rho^2}{\partial x}
\left(2\psi^{*}_{o}\psi_{o}-\psi_{o}\psi^{*}_{e}+\psi^{*}_{o}\psi_{e}\right)
dx\nonumber\\
&+&\frac{g}{4}\int^{\infty}_{-\infty}x\rho^2\left(\psi_{o}\frac{\partial
\psi^{*}_{e}}{\partial x}-
\psi^{*}_{o}\frac{\partial \psi_{e}}{\partial x}+\psi_{e}\frac{\partial
\psi^{*}_{o}}
{\partial x}-\psi^{*}_{e}\frac{\partial \psi_{o}}{\partial x}\right)dx
\label{b}
\eea
\noindent where $H_{ke}$ and $H_{ko}$ are given by,
\bea
H_{ke}=\frac{1}{2}\frac{\partial \psi^{*}_{e}}{\partial x}\frac{\partial
\psi_{e}}{\partial x},\ \ \ \ 
H_{ko}=\frac{1}{2}\frac{\partial \psi^{*}_{o}}{\partial x}\frac{\partial
\psi_{o}}{\partial x}.
\eea
\noindent If we subtract eq. (\ref{b}) from (\ref{a}), then the left hand side
gives $\frac{dI_{2}}{dt}$, while the 2nd terms in the right hand side generate
the potential part of the total Hamiltonian and the last terms cancel out.
Thus, we recover the equation $\frac{dI_{2}}{dt}=H$. Note that none of the
moments $I_{1e}, I_{1o}, I_{2e}, I_{2o}$ satisfy a decoupled equation like
$I_1$ and $I_2$.

\end{document}